\documentclass[letterpaper, 10 pt, conference]{ieeeconf}  
\usepackage{graphicx}
\usepackage{mdframed}
\usepackage{cite}
\usepackage{comment}
\usepackage{placeins}

\IEEEoverridecommandlockouts                              

\title{\LARGE \bf
Exploring the Use of Social Robots to Prepare Children for Radiological Procedures: A Focus Group Study}

\author{Massimiliano Nigro$^{1}$, Andrea Righini$^{2}$ and Micol Spitale$^{1}$
\thanks{$^{1}$Massimiliano Nigro and Micol Spitale are with the Department of Electronics, Information and Bioengineering, Politecnico di Milano, Milan, Italy. Corresponding author's email:
        {\tt\small massimiliano.nigro@polimi.it}}%
\thanks{$^{2}$Andrea Righini is with the Pediatric Radiology and Neuroradiology Department, Children’s Hospital V.Buzzi, Milan, Italy}%
}
\begin{document}

\maketitle
\thispagestyle{empty}
\pagestyle{empty}

\begin{abstract}
When children are anxious or scared, it can be hard for them to stay still or follow instructions during medical procedures, making the process more challenging and affecting procedure results.
This is particularly true for radiological procedures, where long scan times, confined spaces, and loud noises can cause children to move, significantly impacting scan quality.
To this end, sometimes children are sedated, but doctors are constantly seeking alternative non-pharmacological solutions. 
This work aims to explore how social robots could assist in preparing children for radiological procedures. 
We have conducted a focus group discussion with five hospital stakeholders, namely radiographers, paediatricians, and clinical engineers, to explore (i) the context regarding children's preparation for radiological procedures, hence their needs and how children are currently prepared, and (ii) the potential role of social robots in this process. 
The discussion was transcribed and analysed using thematic analysis. 
Among our findings, we identified three potential roles for a social robot in this preparation process: offering infotainment in the waiting room, acting as a guide within the hospital, and assisting radiographers in preparing children for the procedure.
We hope that insights from this study will inform the design of social robots for pediatric healthcare. 
\end{abstract}

\section{Introduction}
Fear and anxiety can make it challenging for children to stay still and cooperate during medical procedures, in turn affecting the procedures' results.
This is especially true in radiological procedures where patient movement can significantly impact the scans' quality, especially in longer procedures like the Magnetic Resonance Imaging (MRI) \cite{zaitsev2015motion}. 
Children often struggle to remain still during scans due to long scan times, the confined and intimidating environment, and the loud noises \cite{bray2022interventions}. The most common solution in pediatric care is sedation \cite{10.1259/bjr/28871143},  especially for children aged 3 to 5 \cite{wachtel2009growth}. However, sedation is far from the perfect solution as the long-term neurological and cognitive effects of sedation and anesthesia remain uncertain \cite{artunduaga2021safety}. As a result, many hospitals are working towards sedation-free radiological procedures.
Some solutions allowed children to practice undergoing the procedure (for example, MRI), helping them understand what to expect and experience similar sensory stimuli. This was achieved through mock scanners, simulated practice, or play-based learning \cite{bray2022interventions}. 
Other interventions focused on educating children about the procedure using educational videos, coloring books, booklets, and storybooks \cite{bray2022interventions}. 
During these interventions, the hospital staff aims to alleviate children's fears—such as lack of information and fear of the unfamiliar environment—to encourage their cooperation.
In the past, there have been previous works with similar objectives that leveraged social robots. 
Logan et al. \cite{logan2019social} introduced a teddy bear-like social robot to hospitalized children. They found that interactions with the robot exhibited greater joyfulness than those with a plush toy.
Ferrari et al. \cite{ferrari2023design} conducted pilot studies to design interactions to help children cope with pain in a hospital setting.
While focusing on an adult population, Boumans et al. \cite{boumans2023social} developed a robot to explain the physical examinations. Their findings indicated that the robot's explanations were perceived as clear, highlighting the potential of social robots in enhancing understanding of medical procedures.
In contrast to other digital solutions, such as mobile applications, videos, or virtual agents, social robots provide engagement with the physical world and enable natural interactions with humans, e.g., through voice, gestures, or touch \cite{deng2019embodiment, spitale2025vita}. 
Overall, robots for children's healthcare settings already reported significant outcomes in technology acceptance, feasibility, enjoyment, engagement, achievement of therapeutic/education goals, pain, and mental health outcomes \cite{triantafyllidis2023social, abbasi2022can, spitale2023using, abbasi2024analysing}.
Therefore, social robots offer an intriguing prospect in preparing children for radiological procedures. 
In this paper, we explore the potential of social robots as a tool to help prepare children for radiological procedures.
To explore how and at what stage of hospitalization a social robot could assist in the preparation for radiological procedures, we conducted a focus group discussion with professionals from the radiology department at the pediatric hospital Vittore Buzzi located in Milan (Italy).
The focus group had two main objectives. 
The first objective was to understand the hospital's context in preparing children; thus identifying their needs (research question, RQ1-a) and examining the current preparation process used by hospital staff to address them (RQ1-b). The second (RQ2) was to explore how social robots could contribute to this process; thus understanding which roles they could fill and which key characteristics they would need in order to fulfil these roles effectively.

We analyze the focus group data using the Thematic Analysis (TA), a qualitative research method. We hope that insights from this study will inform the design of social robots for pediatric healthcare.

\section{Related Works}
Robots have been widely used in children's healthcare settings. 
One of the main applications that robots saw has been distracting children during procedures that require needle insertion \cite{rossi2020emotional,ali2021randomized,rheel2022effect,jibb2018using}. 
By leveraging their expressive capabilities, social robots have been shown to lower children's distress \cite{jibb2018using,rossi2020emotional,ali2021randomized, rheel2022effect}, perception of pain, anxiety, and increase children's happiness \cite{rossi2020emotional}. 
Other works instead focus on improving children's experiences during hospitalization by either helping them cope with pain \cite{ferrari2023design} and stressful situations \cite{crossman2018influence} or entertaining them by providing joyful interactions \cite{logan2019social}. 
In these cases, robots can make the hospital environment feel less intimidating for children \cite{logan2019social}, provide opportunities to boost children's positive moods \cite{crossman2018influence}, and improve their ability to cope with pain \cite{ferrari2023design}.
Finally, social robots have also been used in group therapy sessions for children with cancer, working alongside psychologists to create a safe space where children can express their fears and concerns while learning about their conditions \cite{alemi2016clinical}. These interventions showed that children in groups that leveraged a social robot experienced a significant reduction in stress, depression, and anger compared to the control group.
Boumans et al. \cite{boumans2023social}, in a study with an adult population, demonstrated that a social robot could effectively explain medical tests—specifically, blood pressure and grip strength measurements—to participants in a real-world setting. The participants found the explanations clear, underscoring how promising social robots are for explaining medical procedures effectively.

While these interventions motivate us on the potential of social robots for preparing children for radiological procedures, we still need insights from stakeholders ,like paediatricians and radiographers, to build a succesful and useful robotic system.



\section{Methods}
\begin{table}[htpb]
\caption{Structure of the focus group discussion}
\label{tab:structure}
\begin{tabular}{l l}
\hline
 Item &  Duration \\ 
 \hline
    Discussion on children's needs &  15 min \\ 
    Discussion on current preparation solutions &  15 min \\ 
    Introduction to social robots and demo &  10 min \\ 
    Co-Design activity, filling ``Robot Design Guidelines'' Canva &  10 min \\ 
    Discussion on social robots for preparation &  20 min \\ 
    Conclusion &  5 min \\ 
\hline

\end{tabular}
\end{table}
To gather insights from stakeholders, we chose to organize a \textbf{focus group discussion}. 
This approach has been successfully used in HRI research, such as designing robots for elderly care \cite{soraa2023older}, understanding perceptions of educational robots \cite{smakman2021moral}, or exploring robotic failures during robotic coaching \cite{axelsson2024oh}.
In our focus group, we aim to answer our research questions, namely understand the hospital’s context in preparing children by
identifying their needs (RQ1-a) and examining the current preparation process used by hospital staff to address them (RQ1-b); and explore how social robots could contribute to this process (RQ2). 
To analyze qualitative data from our focus group, we have used the \textbf{Thematic Analysis} (TA). This method has already been employed in HRI research when analysing qualitative data, with applications such as designing robot well-being coaches \cite{axelsson2021participatory, axelsson2024robots}, identifying moral considerations for social robots in education \cite{smakman2021moral}, and analyzing themes in dementia care data \cite{hung2021exploring}. Given its proven effectiveness, TA is well-suited for our study. 

\subsection{Focus Group Study}
During the focus group, we explored how and at which stage a robot could assist in preparing children for radiological procedures. 
The study was approved by one of the regional ethics committees for hospitals in Lombardy, ``Comitato Etico Lombardia I''.
The session, held at the ``Vittore Buzzi'' Pediatric Hospital in Milan, lasted approximately 75 minutes.
To reflect our research questions, the discussion was structured around two main topics as also detailed in Table \ref{tab:structure}:
\begin{enumerate}
    \item The context in the hospital regarding children’s preparation for radiological procedures (children's needs and current methods of preparation)
    \item Potential roles of a social robot in the preparation for radiological procedures.
\end{enumerate}
Between the two discussions, participants were introduced to the social robot Misty II in a demo of around 10 minutes. In the demo Misty was engaging in a small talk conversation with one of the participants.
After that, participants engaged in a 10-minute co-design activity using the ``Design Guidelines'' canvas from \cite{axelsson2021social}. This activity allowed each participant to flesh out their ideas on how and where a robot could be useful in the preparation phase before delving into the discussion. 
One researcher managed the focus group by following a predefined structure, as outlined in Table 
\ref{tab:structure}, to guide the conversation.

\subsection{Participants}
The focus group consisted of five participants (4M:1F): one pediatric radiologist, three radiographers, and one clinical engineer. While we did not collect exact ages, we recorded age ranges. The group included one participant aged 26–35 (ID: P1), one aged 36–45 (ID: P3), one aged 46–55 (ID: P2), and two aged 56–65 (IDs: P0 and P4). On average, participants had 21.5 years of experience working in a hospital setting. 

\subsection{Thematic Analysis}
We employed the Thematic Analysis (TA) to analyze the qualitative data gathered from the focus group discussion. Following the 6-step method outlined by Braun and Clarke \cite{clarke2017thematic}, one researcher used the transcription of the 1-hour and 17-minute focus group discussion, along with the results of the design canvas, as the data corpus. 
The researcher went through three revisions of the data before converging to the final themes presented in the findings.
As the focus group was held in Italian, the quotes reported in the findings section come from the translation of the transcript. The quotes will be reported using participant IDs.

\section{Findings}
\begin{figure*}[htpb!]
    \centering
    \includegraphics[width=\textwidth]{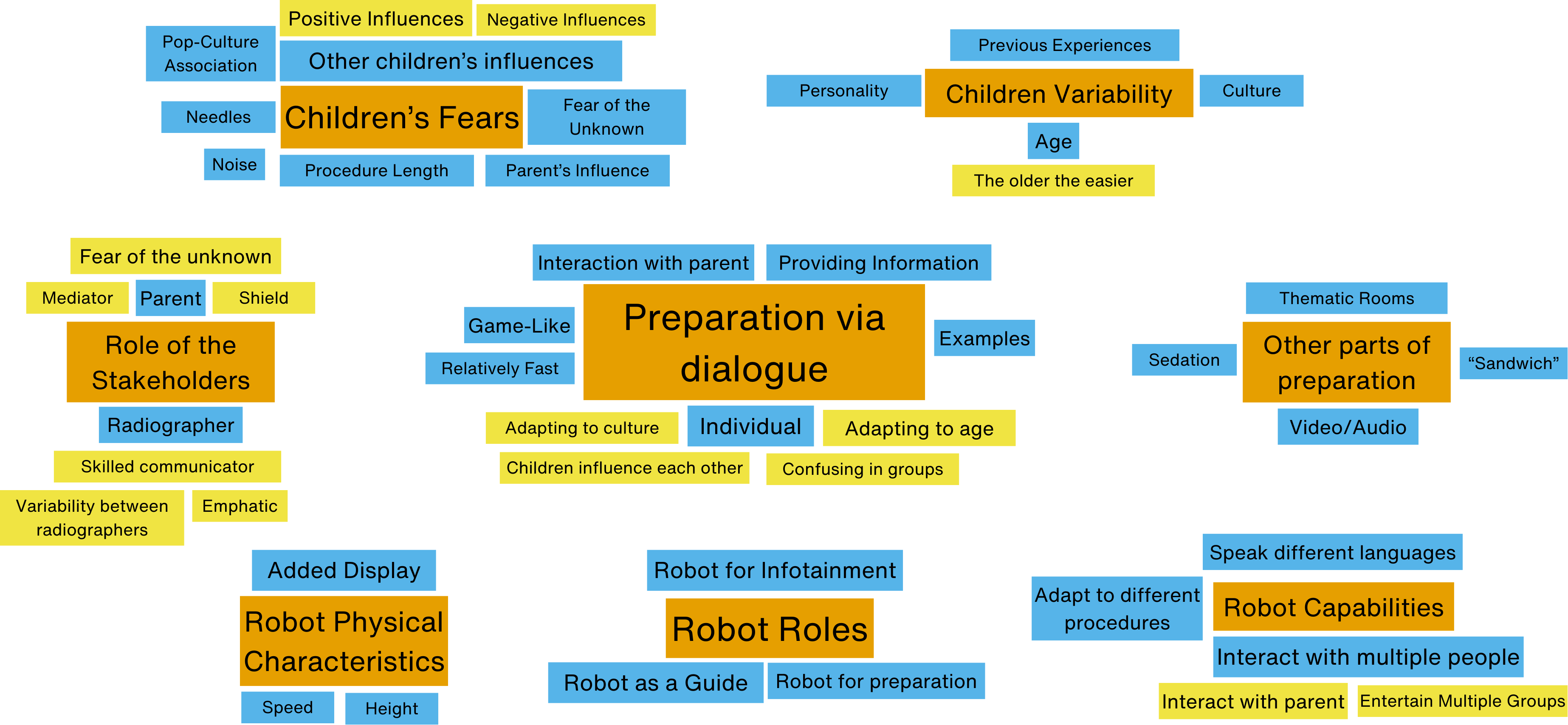}
    \caption{Themes defined in the TA are presented in orange, while codes related to these themes are presented in blue, and sub-codes are presented in yellow}
    \label{fig:themes}
\end{figure*}
Our analysis defined eight themes (counts of occurrences into brackets): children's fears (20), variability (12), role of the stakeholders (15)
, preparation through dialogue (35), other parts of preparation (11), robot roles (28), robot physical characteristics (16), and robot capabilities(12). 
The themes are presented in Figure \ref{fig:themes}.
We present these themes by grouping them into the three main topics of the focus group discussion that aims at addressing our research questions in the following paragraphs. 

\subsection{Children's needs before and during radiological procedures (RQ1-a)}
The \textbf{fears} and feelings of anxiety are mainly what make radiological procedures difficult for \textbf{children}. These sentiments can arise for a variety of reasons. 
The main aspect that causes fear in children is not knowing about the procedure, as highlighted by P0, who mentioned that ``the lack of knowledge, the fear of what they are going to do is one of the main causes of worry and uncertainty in children and therefore the cause of poor cooperation''.
One of the central aspects in the preparation of children for radiological procedures will be explaining the procedure itself to alleviate this fear.
Other fears may arise later during the procedure when hearing intense noises from machines such as the MRI machine, seeing the needle used for procedures that require injections, or being fatigued in more lengthy procedures (such as MRI). 
Another important aspect that emerged is the influence of peers or parents on children's fears. 
Parents can transfer their feelings of fear and anxiety to children, so much so that P1 said, ``[..] you understand whether or not you will be able to take the exam by looking at the parents.'' However, children are also influenced by their peers. 
Sometimes, they can be positive influences; for example, when seeing a younger child go well through the procedure, children often get courage. As P2 specified: ``They see the patient in front of them, who is sometimes much younger, and then they say if he made it maybe I can.''. However, these influences can change case by case as children can also get discouraged and not feel up to par with others who managed to go through the procedure well. P3 claimed that ``the effect that can be proposed sees the younger child cope he feels he can't cope and regresses even more because he wonders why I'm older I can't cope I mean maybe he feels at fault''. \newline 
As such, another important theme that emerged was \textbf{children's variability}.
 Children's approach to radiological procedures changes significantly with age. 
 Younger children are generally more fearful, with children under 6 years old being the most critical. As P0 remarked:``our main target is from a year to 6 years old''. The effect of age on fear alleviates closer to the adolescent age, when carrying out procedures without sedation becomes simpler, as highlighted by P1, who mentioned: ``to a large part of our of children range from zero to ten years, of course in adolescence everything becomes easier''. Consequently, age also affects how radiographers prepare children. An essential step in the preparation is explaining the method with examples the child can relate to and these examples change with age, as children of different ages are interested in different things. As P0 concluded:``Because at that point the dialogue becomes ultra specific, specific to the child's type, age, socio-cultural background including country of origin and age etc. and types of exams''. 
Other factors that change how children approach radiological procedures are personality, differences in culture, and previous experiences. 
Children with different personalities can be more or less influenced by other peers and be more or less courageous in tackling radiological procedures, as also mentioned above. 
Similarly, with culture, children from different cultural backgrounds can cope with fear in various ways, as children coming from stricter educational backgrounds tend to adhere more to the radiographer's instructions and stay still during the procedure. As P0 described: ``Sometimes also the ethnic cultural origin in the sense that paradoxically where there are very strong family institutions very rigid educational systems a little bit in the ancient times the child is more firm, he is intimidated at least''.  
Previous experiences and medical conditions also hugely impact how children tackle the procedure. Children experienced in the procedure will find it easy and be more collaborative. For instance, P1 mentioned that ``[..] there are children who maybe it is the fifth time they come to take the exam are fearless, it is as if they are at home''. 
Even though this can also be dependent on children's personal experience with the exam, as highlighted by P1: ``some children who maybe are oncological but come already done five times the exam come in quietly and get the vein put in, and so I am very reassured and others who are just terrified to have to do the exam again''.

\begin{mdframed}
\textbf{Key Takeaway}: 
The primary source of fear in children undergoing medical procedures is a lack of understanding about what the procedure entails. However, the intensity of this fear varies significantly from child to child. Among the various influencing factors, age plays a particularly important role: younger children, especially those under the age of six, tend to be the most susceptible to heightened fear responses.  
\end{mdframed}
\begin{figure*}[h!t]
    \centering
    \includegraphics[width=\textwidth]{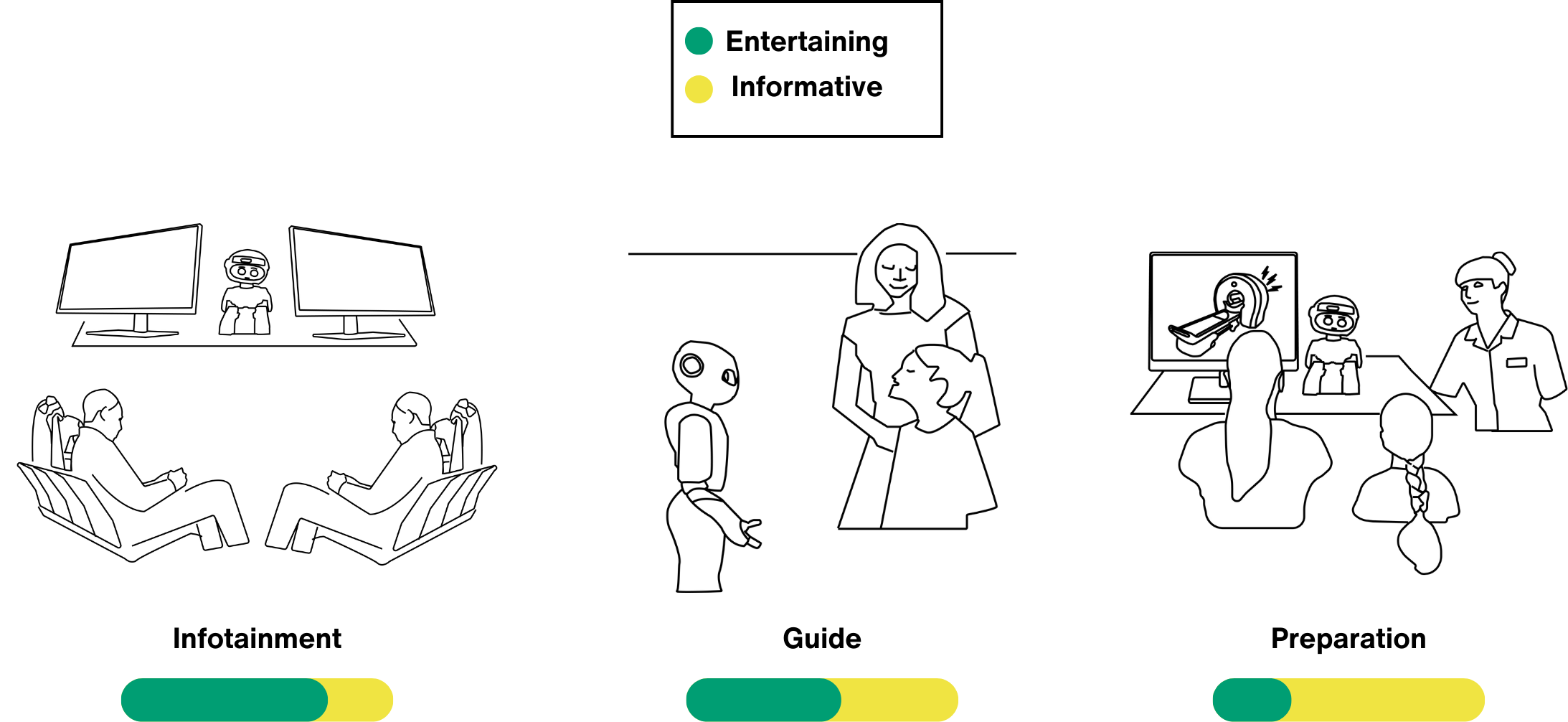}
    \caption{Robot Roles, from left to right: robot for robot for infotainment, robot as a guide, robot for preparation. Yellow indicates how informative the robot should be in the role, and green indicates how entertaining it should be.}
    \label{fig:robot_roles}
\end{figure*}
\subsection{Current methods used to prepare children for radiological procedures (RQ1-b)}
Dialogue with the radiographers is one of the main tools used to prepare children for radiological procedures. 
The \textbf{parent} and the \textbf{radiographer} are the two central figures during this phase. 
Parents often assume the role of mediators between radiographers and children as children, especially of very young age, usually cannot speak very well. P2 pointed out that ``is necessary, for example, maybe a three-year-old speaks but not very well and maybe he cannot understand well so maybe it is the parents who communicate''. Parents also act as a shield for the child, as children are often scared by the unknown environment as highlighted by P2 who described that ``children tend to hide behind their mothers when they are afraid not to approach us''.
Parents can also experience fears and negative feelings, and they can very well transfer those feelings to their children. 
Radiographers, on the other hand, have to navigate these complexities and the variability of children to adequately explain the procedure and ensure that children feel at ease. 
During the focus group, it emerged that they see the ability to communicate effectively as central to preparing children as P1 highlighted ``[..] with the experienced techniques we have in communication, everything becomes faster and more of a game''.
They also emphasized how important it is for the radiographer to be empathetic and adapt to children's sensibilities and emotions. For example, P3 mentioned that ``it always depends on the sensitivity of the child''.
Nevertheless, there is also considerable variability in how radiographers carry out this phase of the preparation both depending on radiographers' skills and workload. As P0 remarked: ``There is a great variability in the operator because there are those who are better at it, who is in a hurry, etc.''. 

Variability aside, some constants emerged on how the radiographers carry out this \textbf{preparation via dialogue}. 
The main objective of this dialogue is to explain to children and parents what the children will have to do during the procedure; however P0 stressed that ``the main thing is to prepare [..] first and foremost the children, for what they are going to do in radiology because knowing what it does, what they are going to do, takes away a lot of the anxiety and uncertainty''.
This is carried out by using examples that the child can relate to. These examples range from tried-out general sentences. For instance, P1 presented an analogy between the MRI room and a spaceship as follows: ``In the spaceship room, there will be noises like your daddy's moped, there will be pillows that keep you still and keep you soft and warm''. P2 also highlighted that these examples are age-dependent and he mentioned that ``if I'm talking to 2-year-olds, I'll give examples, for example, Teletubbies, and if they're 10 or 8 years old, the Teletubbies are outdated, then I can talk about something like football or something... to entertain them and make them feel a bit relaxed''. 
One important aspect that emerged was the importance of keeping this phase playful, making it feel less like an exam and more like a game. P1 for instance explained that he would say something like ``Come on, take the picture you have to smile''.
The participants emphasized the need for this preparation phase to be individual and personalized for each child, as children are highly variable, and interacting with multiple children at the same time would lead to distractions. P1 pointed out that ``single interaction is much more effective when the child knows that he is talking to me and not to three at the same time, in my opinion, he can see much more he is much more attentive'' and P2 again believed that ``doing the same preparation for everyone while they are talking to me as well as others might be a bit confusing''.

Participants also specified \textbf{other factors} that go into the \textbf{preparation} of children, like having a child-friendly hospital environment. P0 mentioned that ``the environment that [needs to be] built in a child-friendly manner to facilitate the approach then architecture, decoration ''. He also suggested that the use of video and audio can help making children at ease like ``multimedia materials that are writings, videos and audios more or less precisely targeted according to the various methods for the age groups''. 
Another technique used by the participants, named the ``Sandwich'', consisted of having the parent present inside the machine with the child. P2 explained that they ``can practically put the child lying down and the parent with his or her head on the child's belly and thus the child can touch [..]''.
If these techniques are ineffective in calming the child, the last resort is sedation. 

\begin{mdframed}
\textbf{Key Takeaway}: 
The most important phase of the preparation is the dialogue with the radiographer. In this dialogue, the radiographer explains the procedure to children and parents playfully, using examples to which the child can relate. This phase is highly personal, and the radiographers adapt to children's age and sensibility; as such, they believe it is best to prepare one child at a time. 
\end{mdframed}

\subsection{Potential roles of a social robot in the preparation for radiological procedures (RQ2)}

As a result of our analysis, we identified three main \textbf{roles} that \textbf{robots} could fill to aid in children's preparation for radiological procedures. 
The roles are represented in Figure \ref{fig:robot_roles}.
In each role, the robot's objectives are either entertaining children while waiting, providing information about the procedure, or both. 

The first role was that of a guide. 
In this role, the robot would meet the child and parent in the welcome room and guide them to the waiting room. In the meanwhile, it could establish a connection with the child by making small talk and at the same time start providing information to the child about the procedure. P1 for example described that ``just as the child arrives, he greets him at the reception desk and says 'come, let's go now, I'll explain to you what you have to do, ... what are you here to do today? And so, what's your name? How old are you?' and he begins to talk a little about the exam to prepare for the exam and what the child will have to do until he is accompanied to the exam room by the receptionist, which we will do. It makes the child calmer, maybe if there are delays, so there is also a bit of a waiting time''.
In this role, the robot should alternate between entertaining the child during the waiting time as suggested by P2 ``also wanting an entertainment where at some point As we said if there is a bit of a wait rather than to make a minimum of conversation with the child'' to explaining the procedure and reassuring the child. P2 pointed also out that it might be important to ``intercalate the preparation for the exam and then do things and have the discussion possibly also brought from the point of view of the actual exam and maybe ask questions to the child you are afraid and so to say but look nothing happens, etc.''. 

The second role we identified was a robot for infotainment in the waiting room. The main objective of the robot in this role is to entertain parents and children in the waiting room; at times, it should also provide some general information about the exam. In this role, the robot should be visible to everyone and potentially attract children and parents. As P2 emphasized: ``since it has to be for entertainment if it has to be something... Something that can be seen by more people'' and then again P0 remarked: ``is to make the table with a screen in the middle of the waiting room could be a nice... That is, if the monitor is big... because then in reality our rooms are not very big it is the child himself who approaches or a mother who...''. 

The last role we identified was that of a robot helping in the preparation dialogue. 
The main objective of the robot in this role is to provide information about the procedure. 
In this role, the robot acts as a personal assistant to the radiographer and carries out part or all of the dialogue with children and parents. P0 envisioned a future in which ``costs are not high'' and imagined that ``every radiographer could have his own robot and carry it around as his personal assistant''. 
This robot would carry out the preparation dialogue as the radiographer would, potentially even leveraging additional monitors to display useful information. P0 explained that ``if the robot is connected to the monitor and gives the instructions' look at the monitor and it gives you the instructions''. As emphasized in the discussion, this robot should be able to adapt its explanations based on the procedure as highlighted by P1.

Participants also specified different \textbf{physical characteristics} for these \textbf{robots}. 
Robots for infotainment or preparation were often associated with using additional displays to either get the attention of more people in the waiting room. P3 suggested that ``instead of just one monitor in the robot, put two or three so that if I put it in the middle of a room, anyone can see it''. P2 also recommended that they can also display more information for preparation and ``show you a film of what the MRI will look like''.
For a guide robot, participants specified height and speed requirements instead.
A guide robot had to be tall enough to be visible so as not to cause accidents when moving in the crowded hospital environment as pointed out by P2. 
Also, the guide robot had to be fast enough when accompanying patients not to cause additional delays for the radiographer, and again P2 highlighted that ``if [he] stand[s] there and ha[s] to wait three minutes for this guy to come and bring [him] the patient it becomes more of a disadvantage''. 

Our analysis also identified different \textbf{robot capabilities} that the participants either needed or wished for. 
Firstly, one of the underlying capabilities a robot would need in these roles is the ability to interact with multiple people simultaneously. In every scenario, the robot is always interacting at least with the parent and the child, and as pointed out by P3 the robot `` will always have two interlocutors at the same time, parent and child''.
This becomes even more prevalent with the robot for infotainment, as highlighted by P2, which has to potentially interact with different families in the waiting room at the same time. 
The ability to speak different languages is also something the participants considered valuable. As children from various cultural backgrounds come into the hospital, it would make them feel more welcome to have a robot speak their native language. P4 mentioned that ``languages like Arabic and Chinese would be very useful [..].The Chinese child is accompanied with instruction in his own language''.
\begin{mdframed}
\textbf{Key Takeaway}: Robots can fulfill multiple roles within a hospital setting. For example, they can serve as \textit{guides}, accompanying children up to the point of the preparation discussion with the radiographer. They can also provide \textit{infotainment}, engaging and informing children and their families while they wait, or \textit{assist the radiographer during the preparation} phase by helping to explain the procedure. 
In each of these roles, the robot helps strike a balance between keeping children and parents engaged during waiting times and delivering important procedural information.
\end{mdframed}

\section{Discussion and Conclusion}



Our findings on the children's needs and current preparation methods echo existing medical literature. As we identified, the literature identifies lack of information as one of the primary challenges for children in hospitals \cite{salmela2009child}. Age is also identified as an essential factor, with statistics confirming that children under 6 years old are more commonly sedated before radiological procedures \cite{wachtel2009growth}. 
Similarly, a significant part of the methods used to prepare children for radiological procedures aim to provide information and explain the procedure \cite{bray2022interventions}. 
These insights can help us create better robotic systems. 
While the specific robot roles we identified are not found in previous HRI literature on children's healthcare, some parallels can be drawn. In hindsight, what we define as ``providing entertainment'' or ``providing information'' has appeared in previous studies in various forms. For example, while Ferrari et al. \cite{ferrari2023design}, Crossmann et al. \cite{crossman2018influence}, and Logan et al. \cite{logan2019social} explored robotic interactions aimed at helping children cope with stress or improving their hospital experience, their interactions effectively provide entertainment to children. On a similar note, Alemi et al.'s robot \cite{alemi2016clinical} functioned in a way that aligns with our concept of a preparation robot, helping to deliver information about the children's condition while addressing their fears. 
Overall, the robot roles we identified with this case study could address broader challenges in pediatric healthcare. The robot as a guide or for infotainment could create a more welcoming hospital environment for children.
The robot for preparation could help explain various medical procedures, reduce children's anxiety, and thus improve medical procedure outcomes. 
In our future research, we aim to involve parents and children in pilot studies to understand better how robots from different roles impact children's fear and anxiety.


\addtolength{\textheight}{-10cm}   



\bibliographystyle{plain}
\bibliography{bibliography}

\begin{thebibliography}{10}

\bibitem{abbasi2022can}
Nida~Itrat Abbasi, Micol Spitale, Joanna Anderson, Tamsin Ford, Peter~B Jones, and Hatice Gunes.
\newblock Can robots help in the evaluation of mental wellbeing in children? an empirical study.
\newblock In {\em 2022 31st IEEE international conference on robot and human interactive communication (RO-MAN)}, pages 1459--1466. IEEE, 2022.

\bibitem{abbasi2024analysing}
Nida~Itrat Abbasi, Micol Spitale, Joanna Anderson, Tamsin Ford, Peter~B Jones, and Hatice Gunes.
\newblock Analysing children’s responses from multiple modalities during robot-assisted assessment of mental wellbeing.
\newblock {\em International Journal of Social Robotics}, 16(5):999--1046, 2024.

\bibitem{alemi2016clinical}
Minoo Alemi, Ashkan Ghanbarzadeh, Ali Meghdari, and Leila~Jafari Moghadam.
\newblock Clinical application of a humanoid robot in pediatric cancer interventions.
\newblock {\em International Journal of Social Robotics}, 8:743--759, 2016.

\bibitem{ali2021randomized}
Samina Ali, Robin Manaloor, Keon Ma, Mithra Sivakumar, Tanya Beran, Shannon~D Scott, Ben Vandermeer, Natasha Beirnes, Timothy~AD Graham, Sarah Curtis, et~al.
\newblock A randomized trial of robot-based distraction to reduce children’s distress and pain during intravenous insertion in the emergency department.
\newblock {\em Canadian Journal of Emergency Medicine}, 23:85--93, 2021.

\bibitem{10.1259/bjr/28871143}
Y~Arlachov and R~H Ganatra.
\newblock Sedation/anaesthesia in paediatric radiology.
\newblock {\em British Journal of Radiology}, 85(1019):e1018--e1031, 01 2014.

\bibitem{artunduaga2021safety}
Maddy Artunduaga, C~Amber Liu, Cara~E Morin, Suraj~D Serai, Unni Udayasankar, Mary-Louise~C Greer, and Michael~S Gee.
\newblock Safety challenges related to the use of sedation and general anesthesia in pediatric patients undergoing magnetic resonance imaging examinations.
\newblock {\em Pediatric radiology}, 51:724--735, 2021.

\bibitem{axelsson2021participatory}
Minja Axelsson, Indu~P Bodala, and Hatice Gunes.
\newblock Participatory design of a robotic mental well-being coach.
\newblock In {\em 2021 30th IEEE International Conference on Robot \& Human Interactive Communication (RO-MAN)}, pages 1081--1088. IEEE, 2021.

\bibitem{axelsson2021social}
Minja Axelsson, Raquel Oliveira, Mattia Racca, and Ville Kyrki.
\newblock Social robot co-design canvases: A participatory design framework.
\newblock {\em ACM Transactions on Human-Robot Interaction (THRI)}, 11(1):1--39, 2021.

\bibitem{axelsson2024oh}
Minja Axelsson, Micol Spitale, and Hatice Gunes.
\newblock " oh, sorry, i think i interrupted you": Designing repair strategies for robotic longitudinal well-being coaching.
\newblock In {\em Proceedings of the 2024 ACM/IEEE International Conference on Human-Robot Interaction}, pages 13--22, 2024.

\bibitem{axelsson2024robots}
Minja Axelsson, Micol Spitale, and Hatice Gunes.
\newblock Robots as mental well-being coaches: Design and ethical recommendations.
\newblock {\em ACM Transactions on Human-Robot Interaction}, 13(2):1--55, 2024.

\bibitem{boumans2023social}
Roel Boumans, Ren{\'e} Melis, Tibor Bosse, and Serge Thill.
\newblock A social robot for explaining medical tests and procedures: An exploratory study in the wild.
\newblock In {\em Companion of the 2023 ACM/IEEE International Conference on Human-Robot Interaction}, pages 263--267, 2023.

\bibitem{bray2022interventions}
Lucy Bray, Lisa Booth, Victoria Gray, Michelle Maden, Jill Thompson, and Holly Saron.
\newblock Interventions and methods to prepare, educate or familiarise children and young people for radiological procedures: a scoping review.
\newblock {\em Insights into Imaging}, 13(1):146, 2022.

\bibitem{clarke2017thematic}
Victoria Clarke and Virginia Braun.
\newblock Thematic analysis.
\newblock {\em The journal of positive psychology}, 12(3):297--298, 2017.

\bibitem{crossman2018influence}
Molly~K Crossman, Alan~E Kazdin, and Elizabeth~R Kitt.
\newblock The influence of a socially assistive robot on mood, anxiety, and arousal in children.
\newblock {\em Professional Psychology: Research and Practice}, 49(1):48, 2018.

\bibitem{deng2019embodiment}
Eric Deng, Bilge Mutlu, Maja~J Mataric, et~al.
\newblock Embodiment in socially interactive robots.
\newblock {\em Foundations and Trends{\textregistered} in Robotics}, 7(4):251--356, 2019.

\bibitem{ferrari2023design}
Oriana~Isabella Ferrari, Feiran Zhang, Ayrton~A Braam, Jules~AM Van~Gurp, Frank Broz, and Emilia~I Barakova.
\newblock Design of child-robot interactions for comfort and distraction from post-operative pain and distress.
\newblock In {\em Companion of the 2023 ACM/IEEE International Conference on Human-Robot Interaction}, pages 686--690, 2023.

\bibitem{hung2021exploring}
Lillian Hung, Mario Gregorio, Jim Mann, Christine Wallsworth, Neil Horne, Annette Berndt, Cindy Liu, Evan Woldum, Andy Au-Yeung, and Habib Chaudhury.
\newblock Exploring the perceptions of people with dementia about the social robot paro in a hospital setting.
\newblock {\em Dementia}, 20(2):485--504, 2021.

\bibitem{jibb2018using}
Lindsay~A Jibb, Kathryn~A Birnie, Paul~C Nathan, Tanya~N Beran, Vanessa Hum, J~Charles Victor, and Jennifer~N Stinson.
\newblock Using the mediport humanoid robot to reduce procedural pain and distress in children with cancer: A pilot randomized controlled trial.
\newblock {\em Pediatric blood \& cancer}, 65(9):e27242, 2018.

\bibitem{logan2019social}
Deirdre~E Logan, Cynthia Breazeal, Matthew~S Goodwin, Sooyeon Jeong, Brianna O’Connell, Duncan Smith-Freedman, James Heathers, and Peter Weinstock.
\newblock Social robots for hospitalized children.
\newblock {\em Pediatrics}, 144(1), 2019.

\bibitem{rheel2022effect}
Emma Rheel, Tine Vervoort, Anneleen Malfliet, Jutte van der Werff~ten Bosch, Sara Debulpaep, Wiert Robberechts, Evelyn Maes, Kenza Mostaqim, Melanie Noel, and Kelly Ickmans.
\newblock The effect of robot-led distraction during needle procedures on pain-related memory bias in children with chronic diseases: A pilot and feasibility study.
\newblock {\em Children}, 9(11):1762, 2022.

\bibitem{rossi2020emotional}
Silvia Rossi, Marwa Larafa, and Martina Ruocco.
\newblock Emotional and behavioural distraction by a social robot for children anxiety reduction during vaccination.
\newblock {\em International Journal of Social Robotics}, 12(3):765--777, 2020.

\bibitem{salmela2009child}
Marja Salmela, Sanna Salanter{\"a}, and Eeva Aronen.
\newblock Child-reported hospital fears in 4 to 6-year-old children.
\newblock {\em Pediatric nursing}, 35(5), 2009.

\bibitem{smakman2021moral}
Matthijs Smakman, Paul Vogt, and Elly~A Konijn.
\newblock Moral considerations on social robots in education: A multi-stakeholder perspective.
\newblock {\em Computers \& Education}, 174:104317, 2021.

\bibitem{soraa2023older}
Roger~Andre S{\o}raa, Gunhild T{\o}ndel, Mark~W Kharas, and J~Artur Serrano.
\newblock What do older adults want from social robots? a qualitative research approach to human-robot interaction (hri) studies.
\newblock {\em International journal of social robotics}, 15(3):411--424, 2023.

\bibitem{spitale2025vita}
Micol Spitale, Minja Axelsson, and Hatice Gunes.
\newblock Vita: A multi-modal llm-based system for longitudinal, autonomous and adaptive robotic mental well-being coaching.
\newblock {\em ACM Transactions on Human-Robot Interaction}, 14(2):1--28, 2025.

\bibitem{spitale2023using}
Micol Spitale, Silvia Silleresi, Franca Garzotto, and Maja~J Matari{\'c}.
\newblock Using socially assistive robots in speech-language therapy for children with language impairments.
\newblock {\em International Journal of Social Robotics}, 15(9):1525--1542, 2023.

\bibitem{triantafyllidis2023social}
Andreas Triantafyllidis, Anastasios Alexiadis, Konstantinos Votis, and Dimitrios Tzovaras.
\newblock Social robot interventions for child healthcare: A systematic review of the literature.
\newblock {\em Computer Methods and Programs in Biomedicine Update}, 3:100108, 2023.

\bibitem{wachtel2009growth}
Ruth~E Wachtel, Franklin Dexter, and Angella~J Dow.
\newblock Growth rates in pediatric diagnostic imaging and sedation.
\newblock {\em Anesthesia \& Analgesia}, 108(5):1616--1621, 2009.

\bibitem{zaitsev2015motion}
Maxim Zaitsev, Julian Maclaren, and Michael Herbst.
\newblock Motion artifacts in mri: A complex problem with many partial solutions.
\newblock {\em Journal of Magnetic Resonance Imaging}, 42(4):887--901, 2015.

\end{thebibliography}
\end{document}